\documentclass[
  journal=pasa
]{cup-journal}
\usepackage{graphicx, subfigure}
\usepackage{amsmath}
\usepackage[nopatch]{microtype}
\usepackage{booktabs}
\usepackage{hyperref}

\newcommand\orcidicon[1]{\href{https://orcid.org/#1}{\mbox{\scalerel*{
\begin{tikzpicture}[yscale=-1,transform shape]
\pic{orcidlogo};
\end{tikzpicture}
}{|}}}}
\title{Perhaps there is no brown dwarf desert?  A study of sub-stellar companions with \textit{Gaia} DR3}

\author{A.~L.~Wallace}
\affiliation{School of Physics and Astronomy, Monash University, Clayton, VIC 3800, Australia}
\alsoaffiliation{Centre for Astrophysics, University of Southern Queensland, C, Darling Heights, QLD 4350, Australia}
\email[A.~L.~Wallace]{Alexander.Wallace@unisq.edu.au}

\author{A.~R.~Casey}
\affiliation{School of Physics and Astronomy, Monash University, Clayton, VIC 3800, Australia}
\alsoaffiliation{Center for Computational Astrophysics, Flatiron Institute, 162 5th Avenue, New York, NY 10010, U.S.A.}


\begin{document}

\begin{abstract}
The brown dwarf desert describes a range of orbital periods (<5\,years) in which fewer brown dwarf-mass companions have been observed around Sun-like stars, when compared to planets and low mass stellar companions.  It is therefore theorised that brown dwarf companions are unlikely to form or remain in this period range.  The \textit{Gaia} space telescope is uniquely sensitive to companions in this period range, making it an ideal tool to conduct a survey of the brown dwarf desert.  In this study, we use Bayesian inference to analyse data from nearby (<200\,pc) Sun-like stars in \textit{Gaia}'s DR3 catalogue, assuming single companions.  From this, we identify 2673 systems (2.41\% of the sample) with possible brown dwarf companions in this period range.  Accounting for observational biases, we find that $10.4^{+0.8}_{-0.6}$\,\% of nearby Sun-like stars have astrometric errors consistent with a brown dwarf-mass companion with a period less than 5\,years, significantly higher than previous studies which reported occurrence rates of <1\,\%.  However, we acknowledge the limitations of DR3 and are unable to make a definitive statement without epoch data.  By simulating epoch data with multiple companions, we find that, while some of the data can be explained by multiple low-mass brown dwarf companions and high-mass planets (>10\,M$_{\mathrm{J}}$), high-mass brown dwarfs (>50\,M$_{\mathrm{J}}$) in this period range are comparatively rare.  Finally, we used our studies of the brown dwarf distribution to predict the number of companions in the brown dwarf desert we can expect to discover in DR4.
\end{abstract}

\section{Introduction}
In close orbits around Solar-type stars, the relative scarcity of brown dwarf companions (13--80\,M$_{\mathrm{J}}$) has been an important topic of debate in studies of planetary system formation.  Radial Velocity (RV) studies have shown that only 0.6\% of systems have brown dwarfs with periods less than 5\,years \citep{lineweaver06}.  The distribution of companion mass appears to follow a power law on both sides of this `brown dwarf desert', suggesting two distinct populations.\\
\noindent There are several hypotheses explaining the brown dwarf desert.  One states that brown dwarfs in close orbits are more likely to migrate inwards and merge with the star as the system forms \citep{armitage02}, however companion migration is still poorly understood.  Other possibilities are that brown dwarfs are ejected early in the system's history by interactions with other companions \citep{Whitworth2018}, or predominantly form at wide separations \citep{jumper13}.  In order to understand the reasons for the brown dwarf desert, knowledge of its location and overall shape is essential.\\
\noindent The brown dwarf desert was first identified by RV  measurements.  However, measurements of companion mass by RV alone are handicapped by degeneracy with orbital inclination.  This can be resolved by combining RV measurements with astrometry from the \textit{Gaia} Space Telescope \citep{unger23,Fitzmaurice24,xiao23}.  \textit{Gaia} is most sensitive to companions on orbital periods comparable to the length of observations.  For DR3, this is $\sim$2.8\,years, which makes \textit{Gaia} an ideal tool to probe the brown dwarf desert.  A recent study \citep{unger23} used astrometric fits from \textit{Gaia}'s non-single star (NSS) catalogue and was able to reclassify 13 companions, previously identified as brown dwarfs, as low-mass stars.  A study presented in \citet{stevenson23} constrained the masses of 12 companions from the NSS catalogue, three of which were in the brown dwarf desert, and 19 brown dwarfs from the DR3 binary\_masses data base.  Additionally, \citet{holl23} constrained orbits in the NSS catalogue and modelled the DR3 epoch data in order to validate substellar companion candidates.  However, due to the current lack of epoch data (available in DR4, expected late 2026) there are many potentially detectable companions which do not yet have orbital solutions in the NSS data base.\\
\noindent An indicator of a companion in DR3 is the renormalised unit weight error (RUWE) which is a measure of how the expected position of the photocentre, based on the fitted parallax and proper motion, differs from observations, relative to systematic errors.  A value of 1 indicates a single star solution, whereas higher values could be indicative of a companion.  In previous studies, this `companion RUWE threshold' has been set at 1.4 for \textit{Gaia} DR2 \citep{lindegren18} but recent studies have set the sky-average threshold at 1.25 for \textit{Gaia} EDR3 \citep{penoyre22}.  In this study, we adopt 1.25 as our detection threshold for all sources which ensures we don't miss any potential companions but comes with the caveat that there may be false positives at the lower mass limits.  The value of RUWE due to a companion is primarily affected by the companion's mass and period, with higher masses and periods close to 2.8\,years producing the highest values.  By modelling this relationship, it is possible to estimate companion properties from the RUWE and fitted track parameters \citep{wallace24}.\\
Our ultimate goal is similar to a recent study presented by \citet{kiefer25} which used the proper motion anomaly from \textit{Gaia} and \textit{Hipparcos} combined with RUWE to conduct a survey of planet hosting stars. While this work is similar, we focus on \textit{Gaia} data alone, due to \textit{Gaia}'s sensitivity to the period range of the brown dwarf desert.  We show how companion properties for our systems can be constrained by \textit{Gaia} alone with the release of DR4.  A comparison with the catalogue presented in \citet{kiefer25} is beyond the scope of this study.\\
\noindent In this study, we analyse a sample of nearby (<200\,pc) Sun-like stars (0.5--1.5\,M$_{\odot}$) with no apparent photometric companions, as indicated by their colour and magnitude.  We perform Bayesian inference on systems with sufficiently high RUWE (>1.25) to determine companion properties and gain an understanding of the companion demographics as a function of mass and period.  Finally, we investigate possible degeneracies in \textit{Gaia} DR3 and explore how epoch astrometry from DR4 will be able to explore the brown dwarf desert to a degree of accuracy never before achieved.
\section{Stellar Sample and Prospects for Brown Dwarf Detection}
\label{sec:sample}
\begin{figure}[t]
    \subfigure[H-R Diagram of Sample]{\label{fig:hr} \includegraphics[width=0.8\linewidth]{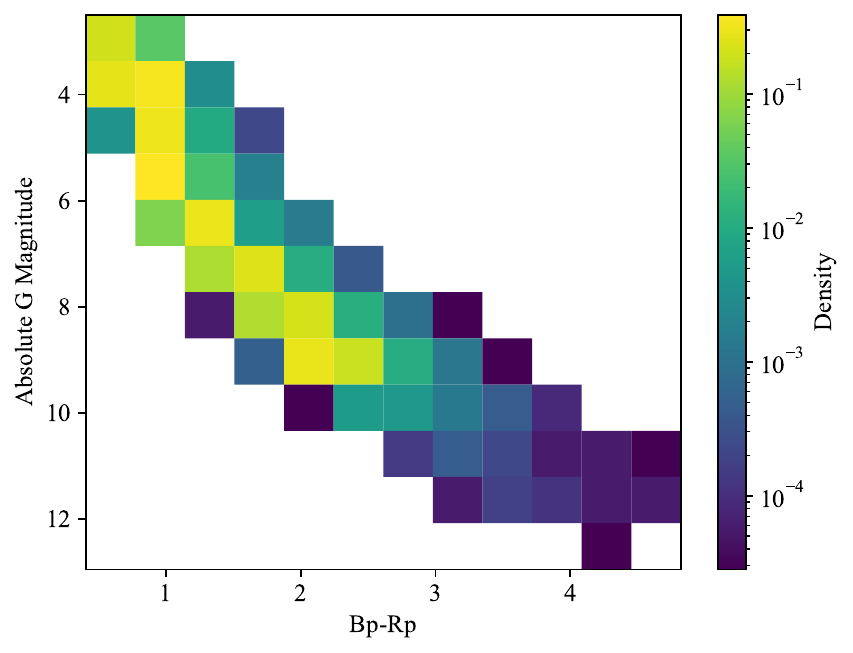}}
    \subfigure[Primary Mass Distribution]{\label{fig:mass_dist} \includegraphics[width=0.8\linewidth]{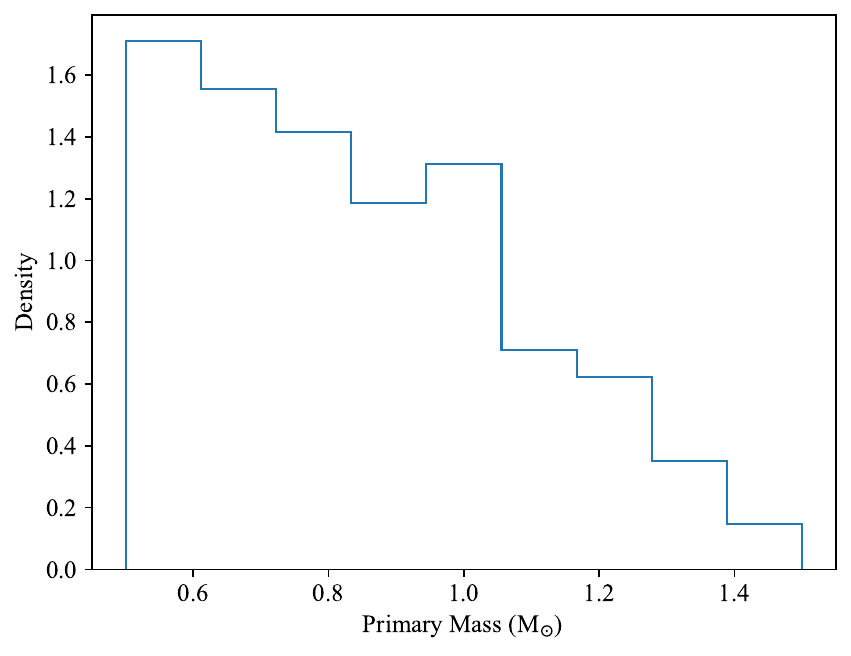}}
    \caption{Colour/magnitude distribution of sample (upper), and calculated mass distribution.  Mass is taken from FLAME models.}
    \label{fig:sample_first}
\end{figure}

\begin{figure}[t]
    \centering
     \subfigure[\textit{Gaia} DR3 94988050769772288]{\label{fig:small} \includegraphics[width=0.8\linewidth]{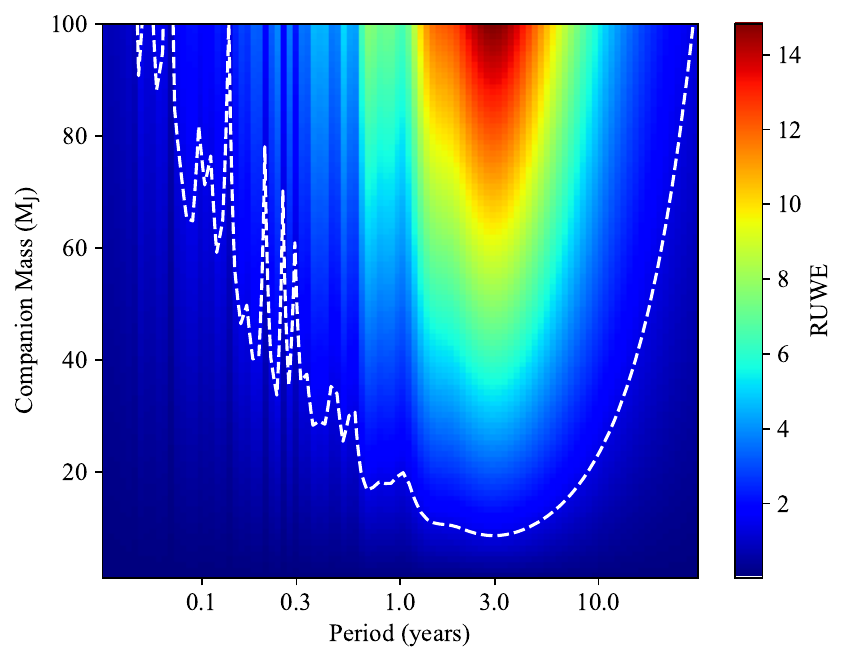}}
    \subfigure[\textit{Gaia} DR3 1985383408935925120]{\label{fig:large} \includegraphics[width=0.8\linewidth]{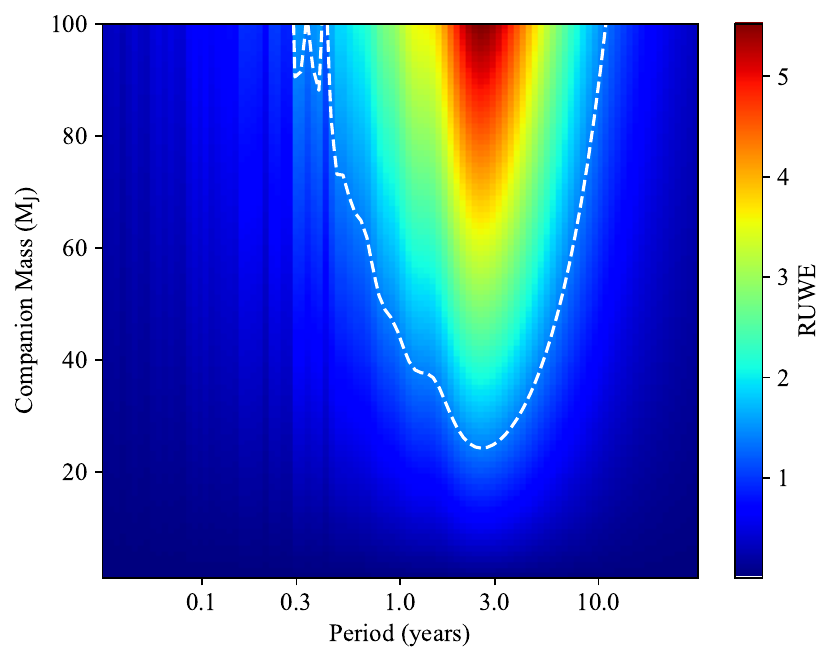}}
    \caption{RUWE as a function of companion mass and age for two examples with estimated masses of 0.75\,M$_{\odot}$ (a) and 1.25\,M$_{\odot}$ (b).  Dashed lines are RUWE from the \textit{Gaia} catalogue.}
    \label{fig:ruwe_ex_both}
\end{figure}

\begin{figure}[t]
    \centering
    \includegraphics[width=1.0\linewidth]{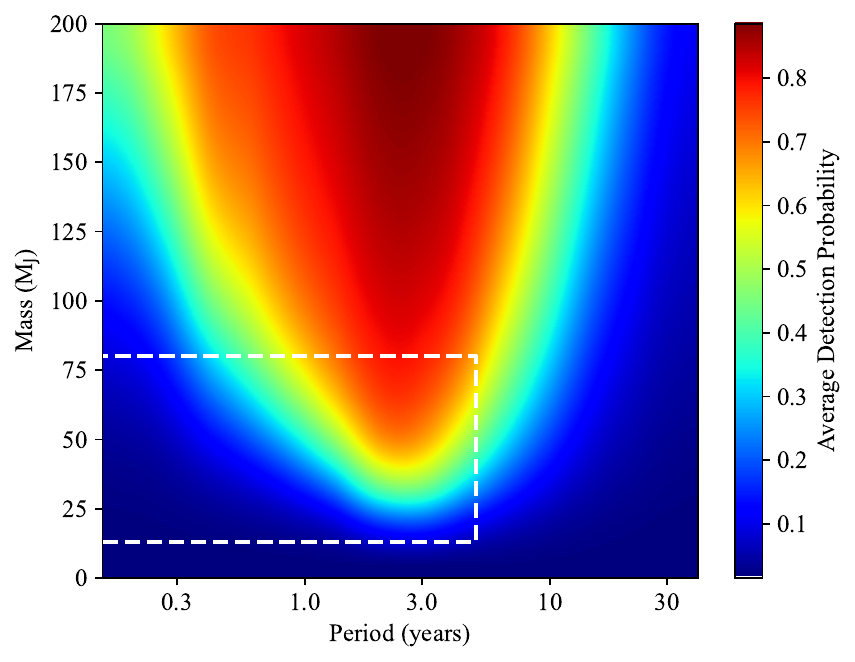}
    \caption{Average detection probability as a function of companion mass and orbital period.  Probability of 1 means the companion is guaranteed to produce a RUWE of more than 1.25 regardless of eccentricity or orbital configuration.  The dashed region marks companions with mass in the brown dwarf range and periods of less than 5\,years.}
    \label{fig:det_prob}
\end{figure}

For this study, a sample of sources in the \textit{Gaia} DR3 main catalogue were selected with RUWE < 20 and parallax > 5\,mas.  The parallax cut was selected to only include nearby stars which are more likely to host detectable brown dwarf mass companions.  The upper RUWE bound was selected to exclude the small number of targets with very clear stellar mass companions or significantly bad astrometric track fitting. A RUWE of 20 is extremely rare and is usually indicative of an extended source or a relatively low number of suitable observations.  In any case, these are unsuitable for our study. This study was also restricted to stars with mass from 0.5--1.5\,M$_{\odot}$, extracted using the FLAME module \citep{kordopatis23}.  We also restricted this to stars on the main sequence, with \texttt{evolstage\_flame} values less than 360.  These cuts resulted in a total sample size of 110,749 stars.  The sample selected on the H-R diagram and a mass histogram is shown in Figure~\ref{fig:sample_first}.\\
\noindent If RUWE is caused by a companion, it is heavily dependent on the companion's mass and period.  For simplicity, this initial study assumes there is only one companion affecting the RUWE.  This is considered a reasonable assumption as we are focusing on main-sequence Sun-like stars, around which very few brown dwarf companions have been found at short periods \citep{Barbato23}.  Although multiple companions have been found on wide orbits \citep{feng22}, we consider the proportion of systems with multiple `desert' companions to be negligible in this initial study.  We also assume the RUWE is caused by a companion.  There may be other explanations, such as a slightly extended source causing an offset in the photocentre but these additional sources of noise are difficult to account for without epoch data.  The derived mass from RUWE could then be taken as a `maximum' mass.  Regardless, a high RUWE could indicate a companion and is helpful in identifying targets for future studies when DR4 is released.  Assuming RUWE is solely caused by a companion, an example of its dependence on a companion's mass and period is shown for two example sources in Figure~\ref{fig:ruwe_ex_both}.  These two examples are taken from the extreme mass ends of the sources shown in Figure~\ref{fig:sample_first}: \textit{Gaia} DR3 94988050769772288 and 1985383408935925120 with mass estimates of 0.75\,M$_{\odot}$ and 1.25\,M$_{\odot}$ respectively and similar RUWE values (1.37 and 1.39 respectively).  The observed RUWE for each is shown on the dashed lines.\\
\noindent The examples shown in Figure~\ref{fig:ruwe_ex_both} demonstrate that, for a range of periods, the RUWE of either source can be attributed to a brown dwarf mass companion, if we assume a circular orbit.  The higher mass target favours a higher mass companion to explain its RUWE value but there are many possibilities in the brown dwarf mass range.\\
\noindent Assuming a companion is detectable if it produces a RUWE>1.25, we can quantify the probability of detecting a companion as a function of mass and period.  To run this study, for each source in the sample, a set of mass and period bins was created.  For each bin, a set of 20 eccentricities, Campbell elements and periastron times were randomly assigned, assuming uniform distributions of $e$, cos$i$, $\Omega$, $\omega$ and $T_{P}$.  This small set of simulated parameters was chosen to minimise computation time and uses the fact that RUWE is only weakly dependent on these parameters when compared to mass or period.  Therefore, we do not need a large number of simulations to get sufficient coverage of RUWE-space.  For each of these sets of parameters, RUWE was calculated and the probability of detecting a companion in the mass-period bin is the proportion which produce a RUWE>1.25.  From this, we have the probability of detection as a function of mass and period.  This calculation ignores the source's RUWE in the \textit{Gaia} catalogue and assumes the companion exists.  The purpose of this calculation is to understand any observational biases which may arise when conducting a statistical study.\\
The average detection probability across all sources is shown in Figure~\ref{fig:det_prob}.\\
Figure~\ref{fig:det_prob} shows a smooth sensitivity map as it is created by averaging over all sources and different orbital configurations.  As a result, there is not such a notable drop in sensitivity at periods of 1 year which arises when an orbit is closely aligned with a star's parallax.  Nonetheless, this still shows a significant increase in detection probability at periods comparable to \textit{Gaia}'s DR3 observing window of $\sim$2.8\,years which is a potential source of bias.
\section{Mass Estimates from Bayesian Inference}
\label{sec:inference}
\begin{figure}[t]
    \centering
    \includegraphics[width=1.0\linewidth]{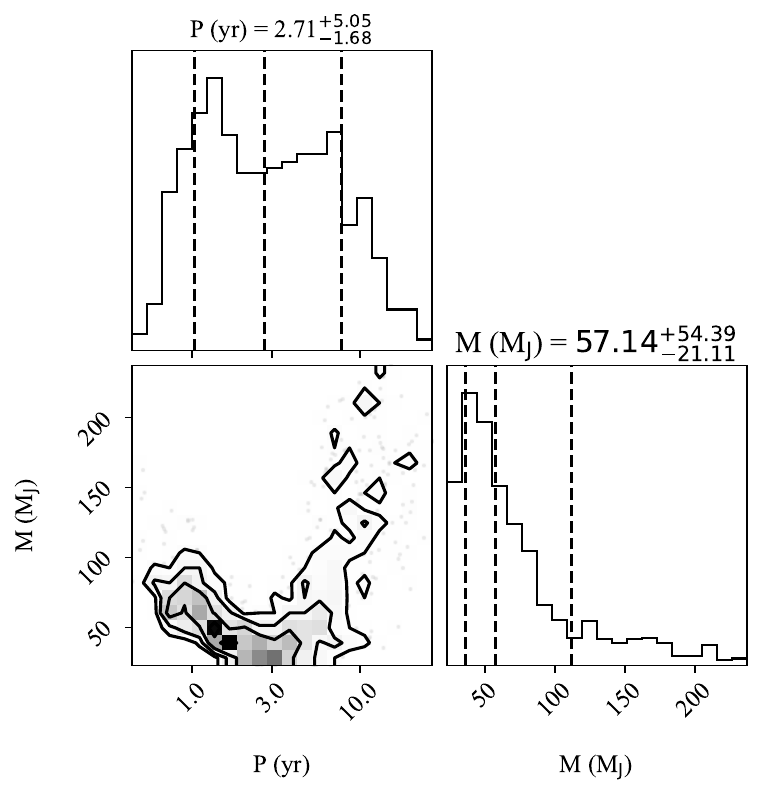}
    \caption{Posterior distributions of inferred mass and period of a companion to \textit{Gaia} DR3 1985383408935925120.}
    \label{fig:corner_ex}
\end{figure}

\begin{figure}[t]
    \centering
     \subfigure[Distribution of mass and periods from all posterior distributions recovered by \texttt{numpyro} inference.]{\label{fig:hist_2d} \includegraphics[width=\linewidth]{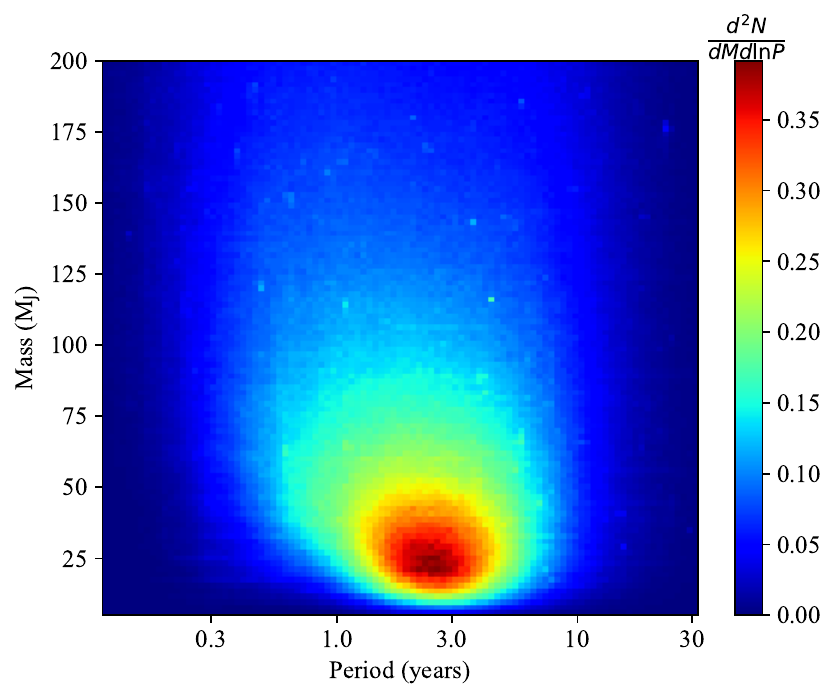}}
    \subfigure[Distribution of periods for brown dwarfs and low mass stars.]{\label{fig:hist_1d} \includegraphics[width=\linewidth]{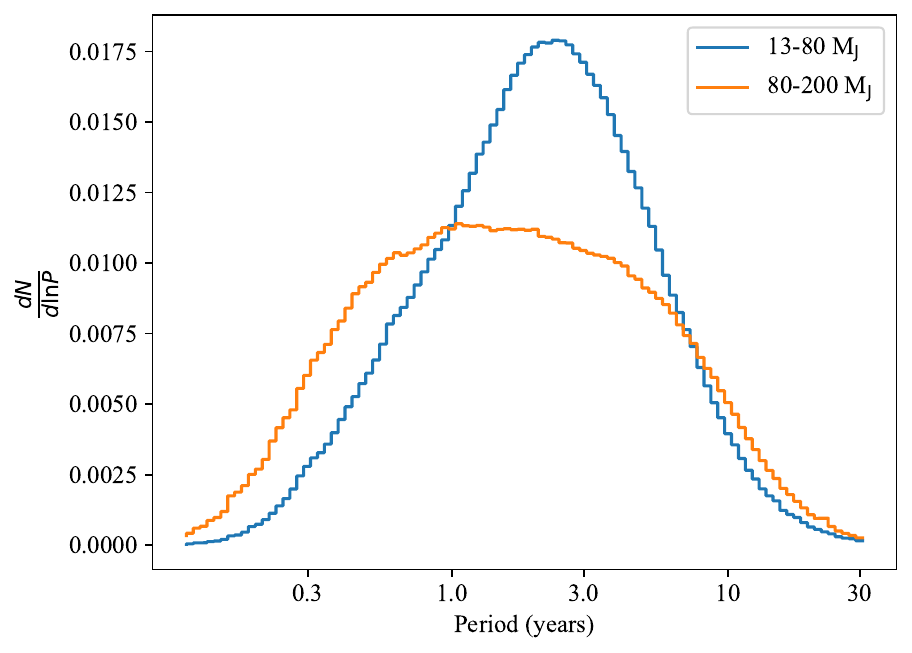}}
    \caption{\textbf{Top: }Distribution of companion masses and periods.  The colour bar is the number of detections per bin divided by the total number of sources (110,749). \textbf{Bottom: }Integrated period distributions for brown dwarf and low-mass stellar companions.}
    \label{fig:mass_sep_hist}
\end{figure}

\begin{figure}[t]
    \centering
    \includegraphics[width=1.0\linewidth]{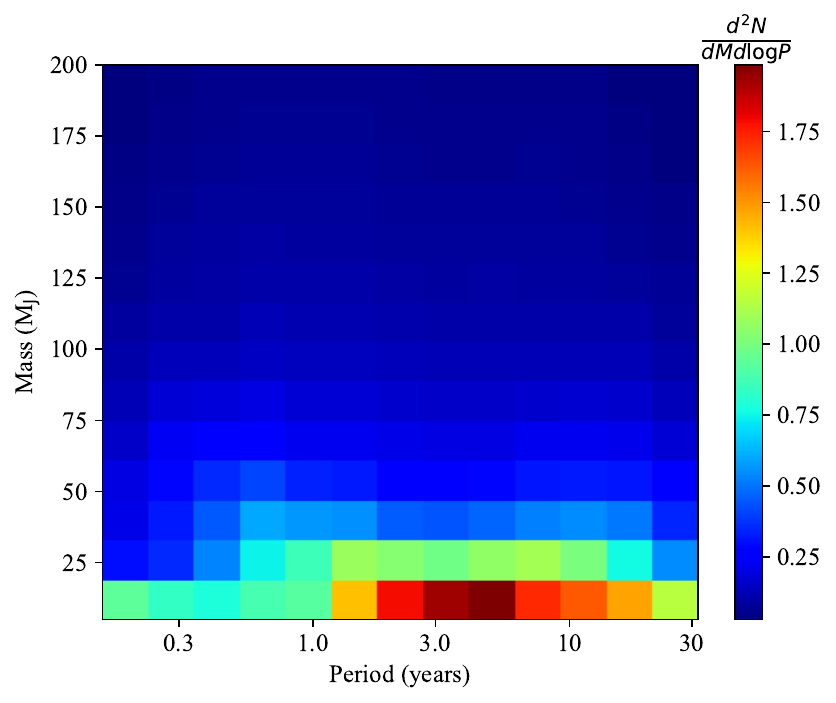}
    \caption{Normalised median occurrence rate as a function of mass and period.}
    \label{fig:med_f_2d}
\end{figure}

\begin{figure}[t]
    \centering
     \subfigure[Occurrence rate of brown dwarfs and stars]{\label{fig:med_f_star} \includegraphics[width=\linewidth]{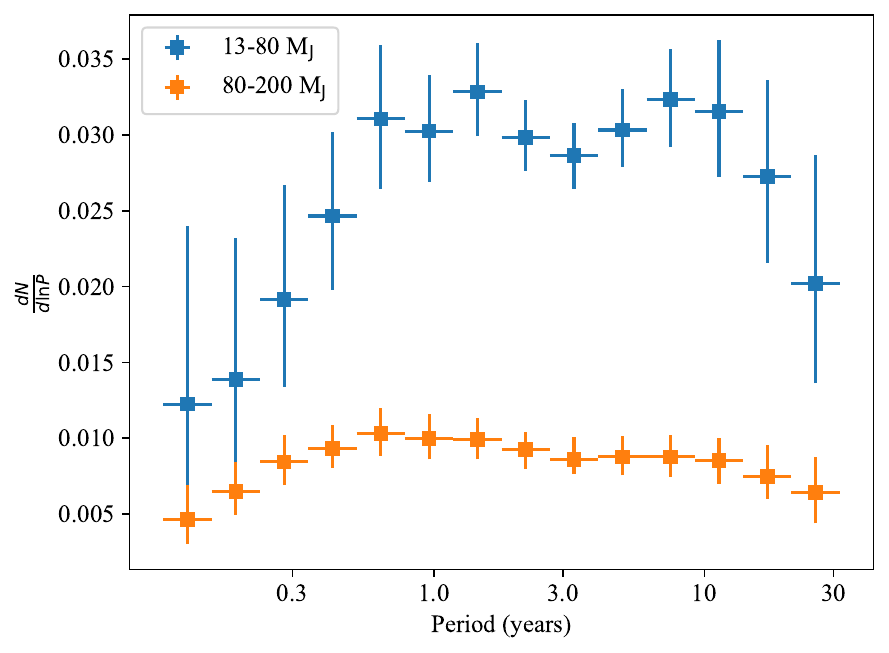}}
    \subfigure[Occurrence rate of two populations of brown dwarfs.]{\label{fig:med_f_bd} \includegraphics[width=\linewidth]{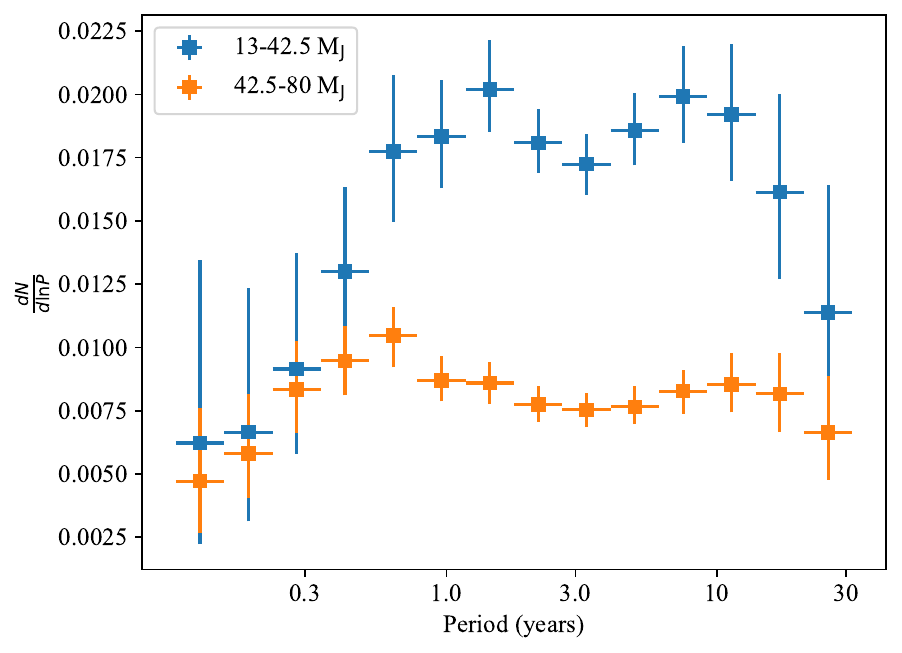}}
    \caption{Occurrence rate as a function of period for brown dwarfs and stellar companions (\textbf{top}) and for low/high mass brown dwarfs as defined by \citet{ma14} (\textbf{bottom})}
    \label{fig:med_f_1d}
\end{figure}

The sources were analysed using the same method as \citet{wallace24}.  This method simulates a star's position as a function of a companion's mass, period and orbital elements, calculates the resultant fitted track parameters (position, parallax and proper motion) in \textit{Gaia} DR3 as well as the RUWE.  This simulates the position of a star along \textit{Gaia's} scanning direction given by:
\begin{equation}
\tilde{x}_{AL} = A\begin{bmatrix}
\Delta\alpha^{*}\\
\Delta\delta\\
\varpi\\
\mu^{*}_{\alpha}\\
\mu_{\delta}\\
\end{bmatrix} + d\tilde{x}_{AL}(M,P,e,i,\Omega,\omega,T_{P}) + \mathcal{N}(\sigma)
\label{eq:pos_single}
\end{equation}
where $\Delta\alpha^{*}$ and $\Delta\delta$ are constant offsets in $\alpha^{*}$ and $\delta$ dimensions, $\varpi$ is the parallax and $\mu^{*}_{\alpha}$ and $\mu_{\delta}$ are the proper motion in $\alpha^{*}$ and $\delta$ dimensions.  This matrix multiplication predicts the position of a single star.  The effect of a companion, given by $d\tilde{x}_{AL}$, and noise are then added.  The matrix $A$ is an $N\times 5$ matrix where $N$ is the number of observation times and is given by:
\begin{equation}
    A = \begin{bmatrix}
\mathrm{sin}\tilde{\psi} & \mathrm{cos}\tilde{\psi} & \tilde{P}_{AL} & \tilde{t}\mathrm{sin}\tilde{\psi} & \tilde{t}\mathrm{cos}\tilde{\psi},
\end{bmatrix}
\label{eq:matrix}
\end{equation}
where $\tilde{P}_{AL}$ is the along-scan parallax factor which depends on the star's sky position as well as the position of the \textit{Gaia} telescope in celestial coordinates $[\tilde{x}_{G},\tilde{y}_{G},\tilde{z}_{G}]$:
\begin{equation}
\begin{split}
    \tilde{P}_{AL} = [\tilde{x}_{G}\sin\alpha-\tilde{y}_{G}\cos\alpha]\sin\tilde{\psi}+\\
    \{[\tilde{x}_{G}\cos\alpha+\tilde{y}_{G}\sin\alpha]\sin\delta-\tilde{z}_{G}\cos\delta\}\cos\tilde{\psi}.
\end{split}
\end{equation}
The observed track parameters are calculated by taking the weighted pseudo-inverse of the matrix $A$:
\begin{equation}
    \begin{bmatrix}
\Delta\alpha^{*}\\
\Delta\delta\\
\varpi\\
\mu^{*}_{\alpha}\\
\mu_{\delta}
\end{bmatrix} = (A^{T}WA)^{-1}A^{T}W\tilde{x}_{AL}
\label{eq:obs_params},
\end{equation}
where $W$ is a weight matrix constructed by comparing the measured position to a single star track and downweighting outlying terms using the Astrometric Global Iterative Solution (AGIS, \citet{lindegren12}) which minimises errors in the solution but cannot eliminate them entirely \citep{penoyre22a}.  The effect of a companion can cause discrepancies between the true and observed track parameters.  For this reason, the observed track parameters are taken as inputs and we solve for the true parameters, in addition to the companion mass and orbital parameters.  The RUWE is calculated by comparing the observed positions $\tilde{x}_{AL}$ with a simulated position $\tilde{x}_{0}$ assuming a single star and applying the matrix multiplication.  RUWE is given by:
\begin{equation}
    RUWE = \sqrt{\sum_{i}^{N} \frac{(x_{AL,i}-x_{0,i})^{2}}{\sigma^{2}(N-5)}},
    \label{eq:uwe}
\end{equation}
The star's RUWE is taken as an additional input and, as shown in Figure~\ref{fig:ruwe_ex_both}, can be linked to the mass and period of a companion.  Thus, using the observed track parameters and RUWE, we apply the equations above to find the most likely companion properties and true track parameters to produce the observed data.\\
While \citet{wallace24} used \texttt{pystan}, this study was implemented in \texttt{numpyro} \citep{phan2019composable} using JAX \citep{jax2018github} for gradient computation and accelerated sampling with the the No-U-Turn Sampler (NUTS) \citep{Hoffman11}.  This proved more efficient in the computationally intense parts of the calculations, namely solving Kepler's equation for the eccentric anomaly.\\
Some examples showing the effectiveness of this method are shown in Figure~\ref{fig:test_cases} in~\ref{sec:test_inference} using simulated companions to real sources in the sample.  The 1-$\sigma$ levels and injected values are shown for comparison.  This test showed that, while we can accurately recover the periods and mass within 1-$\sigma$, this covers a very broad range of mass and period.  We are more accurate for the companions with shorter periods, however, for an example with an injected period of 4\,years, our inference produces a most likely period of 2.8\,years.  This shows we favour solutions close to where \textit{Gaia} is the most sensitive and should be taken into consideration when computing occurrence rates.\\
Applying this to real data, an example corner plot from this method is shown in Figure~\ref{fig:corner_ex} for \newline\textit{Gaia} DR3 1985383408935925120 demonstrating the posteriors on companion mass and period.
The posterior on mass shows a median in the brown dwarf range  but a long tail into low-mass stars.  Due to the low RUWE, there is significant spread in the period posterior though a slight preference toward a short period brown dwarf.\\
Due to the computationally intensive nature, this analysis is only run on sources with RUWE>1.25 and parallax>10\,mas.  This restricts the inference to 3065 sources.  For other sources with RUWE>1.25, the mass of a companion is inferred using the result from a nearby `reference' source with similar coordinates and correcting for the distant source's mass, RUWE and astrometric error.  This approximates the mass distribution of the inferred companion for the more distance source.  The period distribution is the same as for the reference source since, when using RUWE alone to infer these properties, the inferred period distribution is mostly dependent on the observation times and scanning angles, which are only dependent on the coordinates.  An example showing the utility of this method is shown in Figure~\ref{fig:ref_ex} in~\ref{sec:ref_ex}.
For sources with RUWE<1.25, no companions are inferred.\\
\noindent Figure~\ref{fig:mass_sep_hist} shows the distribution of companion masses and periods from this analysis as well as integrated period distributions for brown dwarf and stellar companions.  The distribution was calculated by combining the posterior distributions of all targets.\\
\noindent The distribution in Figure~\ref{fig:hist_2d} shows a majority of detections with periods between 1 and 3\,years, which is due to \textit{Gaia}'s detection sensitivity, shown in Figure~\ref{fig:det_prob}.  The brown dwarf and low-mass star period distributions shown in Figure~\ref{fig:hist_1d} exhibit similar behaviour, though the distribution peak at $\sim$2.8\,years is slightly more pronounced for brown dwarfs.  This is most likely due to their detectability having a higher dependence on period, on account of their lower mass.  In total, we found 2673 systems in which $>68$\% of the posterior indicates a brown dwarf companion with periods less than 5\,years, giving an initial occurrence rate of 2.41\%.  However, the period distributions in Figure~\ref{fig:hist_1d} are highly affected by \textit{Gaia}'s detection sensitivity and are most likely not indicative of the real distribution.\\
In order to calculate the distribution of companion masses and periods, while taking observational biases into account, we use the methods from \citet{lafreniere07} which had been previously applied to giant planet statistics \citep{vigan2012international,wallace2020high}.  In this method, a likelihood $\mathcal{L}$ of data $d$ given an occurrence rate $f$ (in a particular mass-separation bin) is given by:
\begin{equation}
    \mathcal{L}(\textbf{d}|f) = \prod\limits_{i=1}^{N}(1-fp_{i})^{1-d_{i}}(fp_{i})^{d_{i}}
\end{equation}
where $N$ is the number of sources (37,231 in this case), $f$ is the fraction of sources with a companion in this mass-separation range, and $p_{i}$ is the probability of detecting a companion around the $i$-th source in a particular mass-separation bin, assuming the companion exists (the average of this is shown in Figure~\ref{fig:det_prob}.  The value of $d_{i}$ is either 1 or 0, indicating a detection or non-detection respectively (the average of this is shown in Figure~\ref{fig:hist_2d}).  Using Bayes' Theorem, the probability of $f$ given $d$ is given by:
\begin{equation}
    p(f|\textbf{d}) = \frac{\mathcal{L}(\textbf{d}|f)p(f)}{\int\limits_{0}^{1}\mathcal{L}(\textbf{d}|f)p(f)df}
\end{equation}
where $p(f)$ is the prior for the occurrence rate $f$.  We are assuming no prior knowledge of the true companion distribution.  For this reason, we assume the prior is uniform.  Using non uniform priors, if the shape of the prior depends on mass and period, would cause our results to favour certain periods and masses and could be a useful study as more statistical analyses become available.  However, such a study is beyond the scope of this work.  Our probability is then simply equivalent to the normalised likelihood.\\
\noindent From this probability, we can calculate the median occurrence rate, given by the value of $f$ below which 50\% of the probability integral lies.  This median occurrence rate is shown in Figure~\ref{fig:med_f_2d}, where it is normalised by dividing by the size of the mass and log(period) bins.\\
\noindent The median occurrence rate demonstrates an overall increase towards periods of 3--10\,years. Some of this is due to the high number of possible detections around 3\,years shown in Figure~\ref{fig:hist_2d}.  However, the peak extends to longer periods, possibly due to lower detection probabilities and a small but non-zero number of detections which drives up the estimated occurrence rate.  If we extract two populations based on mass we can gain a clearer understanding of the brown dwarf desert.  We have done this first comparing brown dwarfs (13--80\,M$_{\mathrm{J}}$) and low-mass stars (>80\,M$_{\mathrm{J}}$), and then two distinct populations of brown dwarfs, below and above 42.5\,M$_{\mathrm{J}}$ as defined by \citet{ma14}.  Both of these are shown in Figure~\ref{fig:med_f_1d}.\\
\noindent The occurrence rates in Figure~\ref{fig:med_f_star} demonstrate an increase towards longer periods for brown dwarfs up to periods of $\sim$1\,year, after which there is little variation with period.  This behaviour is repeated when we compare the two brown dwarf populations in Figure~\ref{fig:med_f_bd}.  This is further evidence that brown dwarfs at short periods are comparatively rare, however we note that similar behaviour is observed with low mass stars.  Overall, these results suggest the brown dwarf desert isn't as dry as previous studies have suggested: integrating over period, we find $10.4^{+0.8}_{-0.6}$\,\% of sources have a plausible brown dwarf companion with period less than 5\,years, while $4.6^{+0.5}_{-0.4}$\,\% have a brown dwarf companion with period greater than 5\,years.  However, we acknowledge our method is inaccurate for long periods so more analysis (probably with the upcoming DR4) is required to make a definitive statement on this.  Among the short period brown dwarfs, we found that $6.0^{+0.5}_{-0.3}$\% have a companion below 42.5\,M$_{\mathrm{J}}$ and $3.2\pm 0.2$\,\% have a companion above 42.5\,M$_{\mathrm{J}}$.  This, however assumes a RUWE threshold of 1.25 which could allow for false positives.  If we set the threshold at 1.4, as was the case for DR2, we get a brown dwarf occurrence rate of $4.9^{+0.6}_{-0.3}$\,\% with periods less than 5\,years, which we can take as our lower estimate.  Even with this lower estimate, our brown dwarf occurrence rates are significantly higher than the value reported in \citet{lineweaver06} of <1\,\%.  This could be due to previous surveys focusing on RV data which is limited to minimum mass calculations given by Msin$i$.  Astrometry can resolve this degeneracy and it is possible that a significant number of planetary companions can be reclassified as brown dwarfs, as suggested by some results in \citet{kiefer21} and \citet{wallace24}.  A higher brown dwarf fraction at small separations has implications for formation models.  It is possible that disc fragmentation could occur at smaller separations than previously thought \citep{Vorobyov13}, or inward migration occurs in a higher percentage of systems \citep{Beuther14}.\\
\noindent The brown dwarf distribution shows a decrease for periods greater than 10\,years while the stellar distribution doesn't show as steep a decline.  This could be due to a lack of disk material at wide separations making it difficult to form brown dwarfs and planets, whereas stellar companions could be captured at these distances from other systems.  These results indicate that the brown dwarf desert may not be as dry as originally observed as previous studies were unable to effectively probe \textit{Gaia}'s period range.  However, without access to the epoch data, it is impossible to distinguish one high mass companion from multiple low mass companions.  This will become available in \textit{Gaia} DR4 (expected in 2026).
\section{Effect of Multiple Companions}
\begin{figure}[t]
    \centering
    \includegraphics[width=1.0\linewidth]{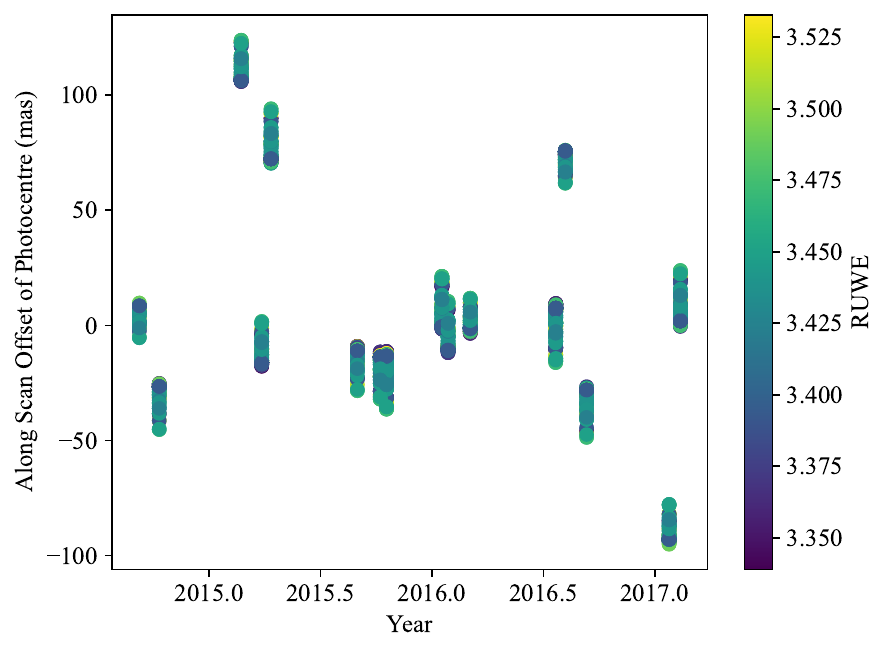}
    \caption{Simulated epoch data for different two companion systems around Gaia DR3 146740207663868032.  The offset is relative to the photocentre position at the DR3 epoch time of 2016.0 in the scanning direction. }
    \label{fig:epoch_ruwe}
\end{figure}

\begin{figure}[t]
    \centering
     \subfigure[Mass as a function of period for given RUWE assuming 1 companion]{\label{fig:m_p_1_comp} \includegraphics[width=\linewidth]{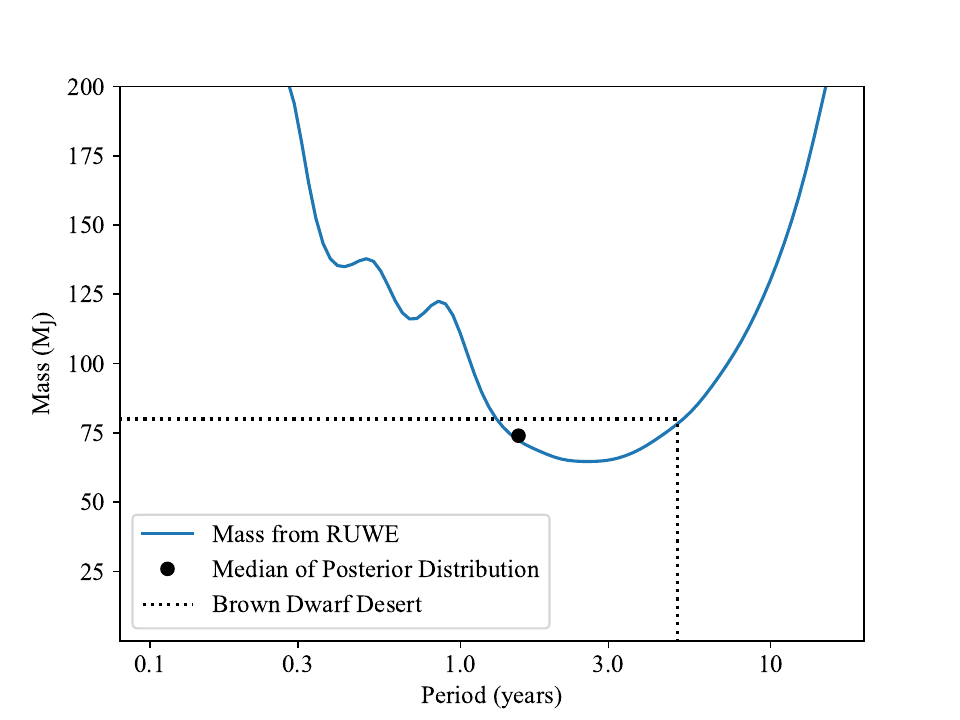}}
    \subfigure[Mass as a function of period for given RUWE assuming 2 companions.  Different colours indicate different cases.]{\label{fig:m_p_2_comp} \includegraphics[width=\linewidth]{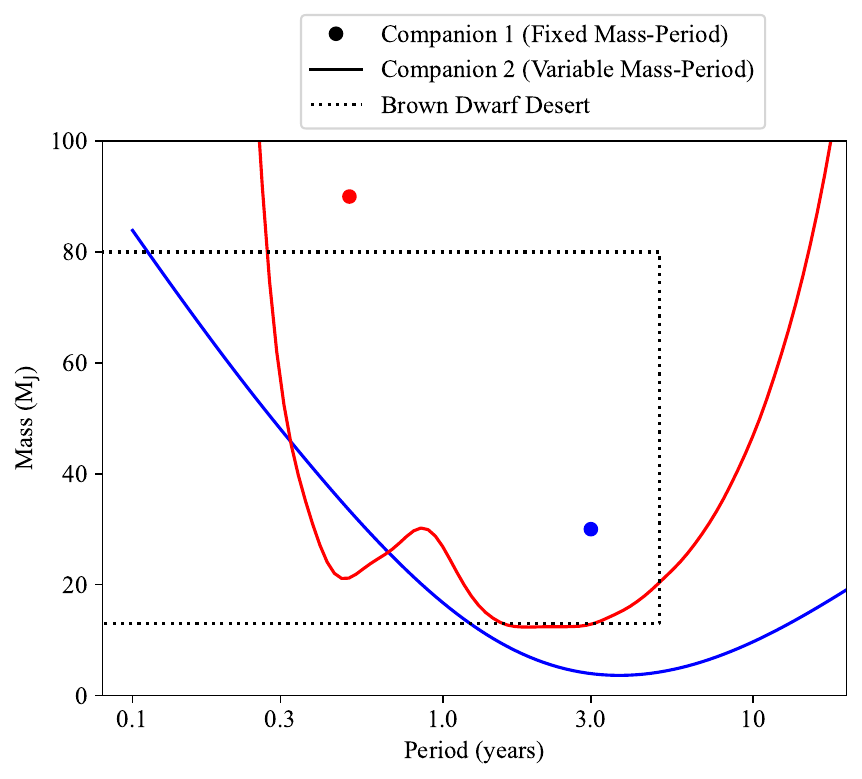}}
    \caption{\textbf{Top: }Mass-period relation for a given RUWE assuming 1 companion, with brown dwarf desert and median of the posterior distribution shown. \textbf{Bottom: }Mass-period relation of a second companion, assuming a specific mass and period of first companion.  Two multi-companion cases are shown for this source.}
    \label{fig:multi_ex}
\end{figure}

\begin{figure}[t]
    \centering
    \includegraphics[width=1.0\linewidth]{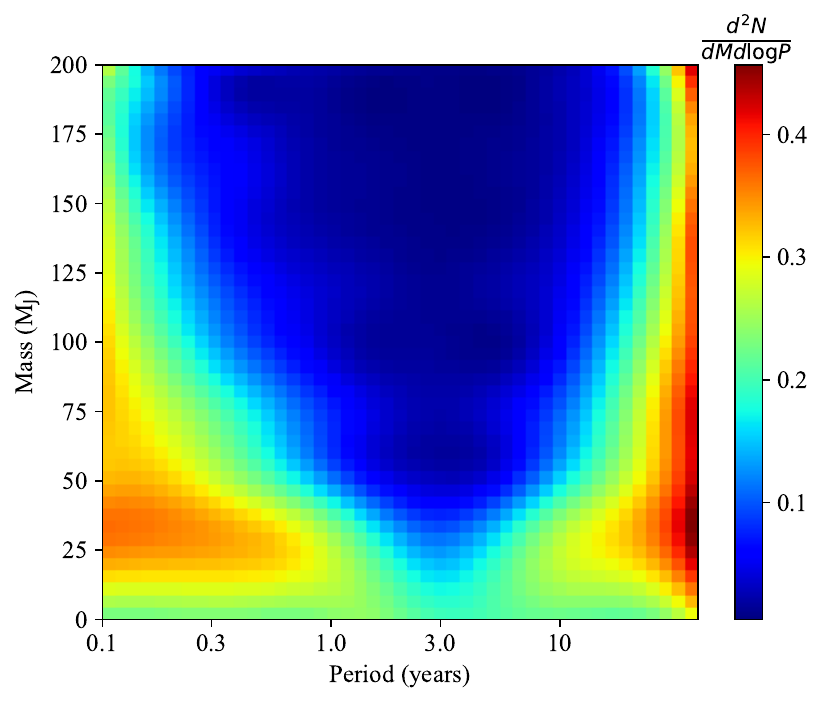}
    \caption{Distribution of all masses and periods assuming two companions.  These were calculated by finding the mass-period combinations for two companions which produced a RUWE within 0.1 of the observed value.}
    \label{fig:dist_2_comp}
\end{figure}

While we can't infer the presence of multiple companions with DR3, it is possible to simulate the effect of a multiple companion system on RUWE and how this differs from the case with a single companion.  In Section~\ref{sec:inference}, we found many sources with RUWE indicative of companions in the brown dwarf desert.  However, there is a wide range of possible multi-companion scenarios which produce the same observed track and RUWE and avoid the brown dwarf desert.  This was tested on the sample from Sections~\ref{sec:sample} and~\ref{sec:inference}, where a set of masses from 0--0.2\,M$_{\odot}$ and periods from 0.1--40\,years were simulated for two companions for each source.  The resultant RUWE values were calculated and the sets of companion masses and periods were selected if they produced a RUWE within 0.1 of the observed value.\\
\noindent For an example source of Gaia DR3 146740207663868032, with RUWE of 3.43, we had previously inferred a companion mass of 74$^{+58}_{-28}$\,M$_{\mathrm{J}}$ with a period of 1.54$^{+3.05}_{-0.87}$\,years, assuming one companion, which puts this most likely in the upper mass range of the brown dwarf desert.  However, if we allow for more than one companion, there are many extra possible configurations.  Each of these configurations produces different epoch data but the same RUWE; a clear limitation of DR3.
\noindent This limitation of DR3 is emphasised in Figure~\ref{fig:epoch_ruwe} which shows the epoch data for possible two companion systems and the distribution of RUWEs these systems create.  For this example, a four-dimensional grid of masses below 200\,M$_{\mathrm{J}}$ and periods below 40\,years was simulated for two companions.\\
\noindent Out of our $30\times30\times30\times30$ grid of companion mass and periods, there are 134 solutions with RUWE within 0.1 of the measured value.  The high variance of the epoch data shown in Figure~\ref{fig:epoch_ruwe} for a limited range of RUWE values shows the degeneracies present in DR3 which will be solved in DR4.  Prior to the release of DR4, we have investigated possible two companion configurations of these sources. For simplicity, inclination and eccentricity are set to 0 and only mass and period are varied.  Using the example of Gaia DR3 146740207663868032, Figure~\ref{fig:multi_ex} shows how the mass varies as a function of period for this source, assuming one companion in Figure~\ref{fig:m_p_1_comp} and two companions in Figure~\ref{fig:m_p_2_comp}.  For the case with two companions, one is given a fixed mass and period (shown with a dot) and the second has mass as a function of period for the given RUWE and taking the first companion into account (shown with a curve).  Two possible configurations are shown in Figure~\ref{fig:m_p_2_comp}.\\
As shown in Figure~\ref{fig:m_p_1_comp}, if we assume a single companion, the source's high RUWE excludes the possibility of a planetary mass companion and is more likely attributed to a high mass brown dwarf or low mass star.  However, as shown in Figure~\ref{fig:m_p_2_comp}, if we assume a first companion in the brown dwarf desert (30\,M$_{\mathrm{J}}$ at 3\,years), there is a chance the second companion is a planet with period >1\,year.  Additionally, if the first companion is a low mass star (0.09\,M$_{\odot}$ at 0.5\,years), the second companion  likely resides in the brown dwarf desert.\\
\noindent Extending this to the entire sample, we can produce a distribution of masses and periods, similar to Figure~\ref{fig:mass_sep_hist} but assuming two companions.  Using the same method as for Gaia DR3 146740207663868032, we simulated the possible mass and period combinations producing the observed RUWE (within 0.1), assuming two companions, for each source.  The total distribution of masses and periods of both companions for all sources is shown in Figure~\ref{fig:dist_2_comp}.
The distribution shows a lack of companions at periods close to 2.8\,years and at high mass, in stark contrast to the probability plot in Figure~\ref{fig:det_prob} and the result from inference shown in Figure~\ref{fig:mass_sep_hist}.  This is most likely due to companions in this mass-period range producing high RUWEs which are amplified by a second companion.  Due to the relative lack of sources with high RUWE, and our own cut on the sample (RUWE<20) this absence is to be expected.\\
The significant difference between this result and the result shown for single companions highlights the limitations of DR3.  Since these masses and periods were calculated based on a preset simulation and not Bayesian Inference, a statistical study similar to Section~\ref{sec:inference} was not attempted for this sample.  An efficient study of the brown dwarf desert which allows for multiple companions will not be possible until DR4.\\
\section{Expected Brown Dwarf Yield from DR4}
\begin{figure}[t]
    \centering
    \includegraphics[width=1.0\linewidth]{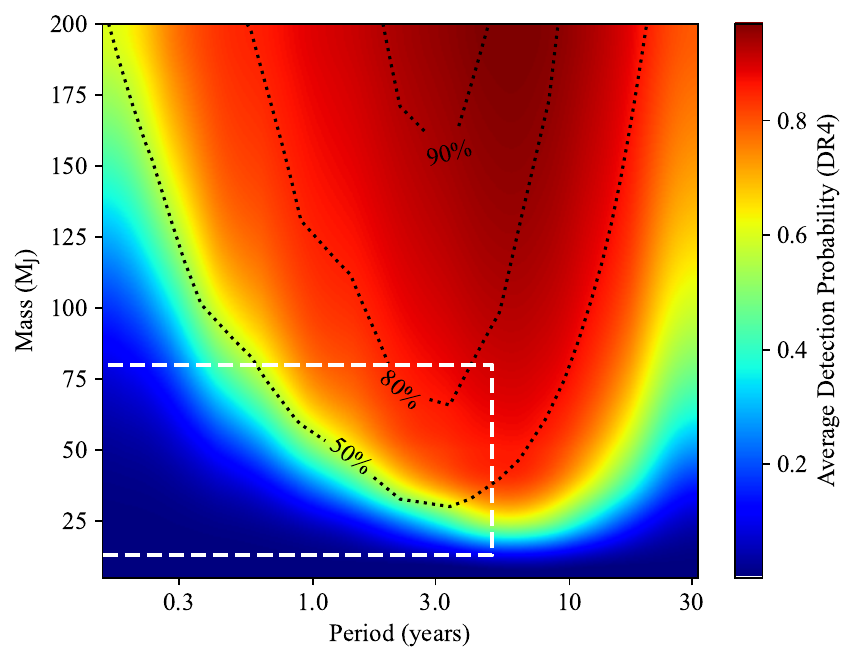}
    \caption{Average detection probability as a function of companion mass and orbital period for DR4 time series.  The white dashed region marks companions with mass in the brown dwarf range and periods of less than 5\,years.  The DR3 detection probabilities of 50\%, 80\% and 90\% are shown with the dotted lines for comparison.}
    \label{fig:det_prob_dr4}
\end{figure}

As shown in Figure~\ref{fig:epoch_ruwe}, different orbital configurations can produce significantly different epoch astrometry with the same RUWE.  This shows great promise for the results of DR4.  This new data release will also include twice as many observations, spanning 66\,months.  Figure~\ref{fig:det_prob_dr4} shows the average companion detection probability for targets in our sample as a function of companion mass and period, similar to Figure~\ref{fig:det_prob}, but with the full time range of DR4.  Probability contours for DR3 are shown for comparison.
Figure~\ref{fig:det_prob_dr4} demonstrates an increase in detection probability for DR4, due to the additional data with greater sensitivity at longer periods.  DR4 is also expected to be capable of probing lower masses and longer periods.\\

\begin{figure}[t]
    \centering
     \subfigure[Brown Dwarf yield assuming distribution from \citet{lineweaver06}.]{\label{fig:bd_det_lit} \includegraphics[width=\linewidth]{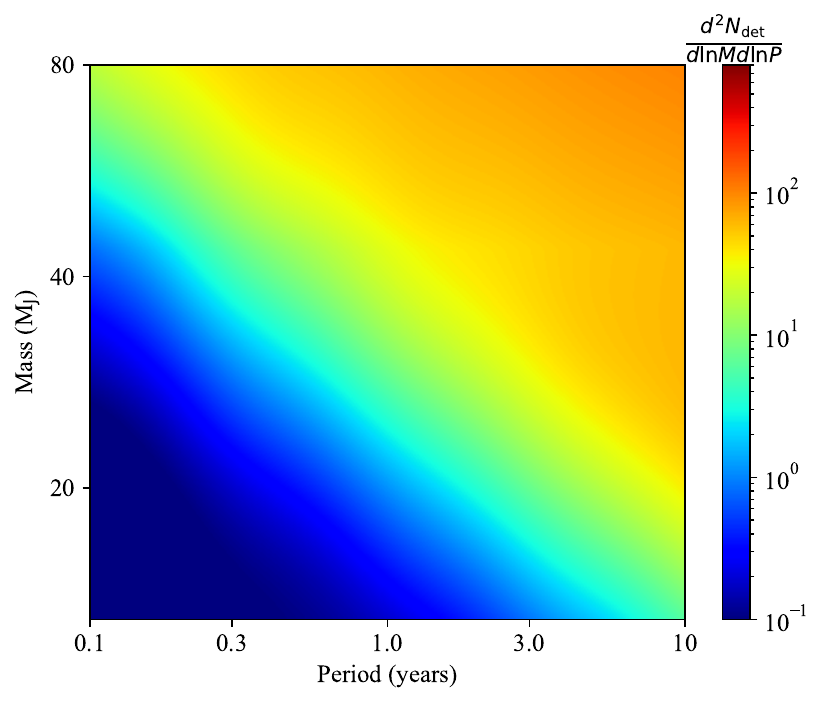}}
    \subfigure[Brown Dwarf yield assuming distribution from this study.]{\label{fig:bd_det_calc} \includegraphics[width=\linewidth]{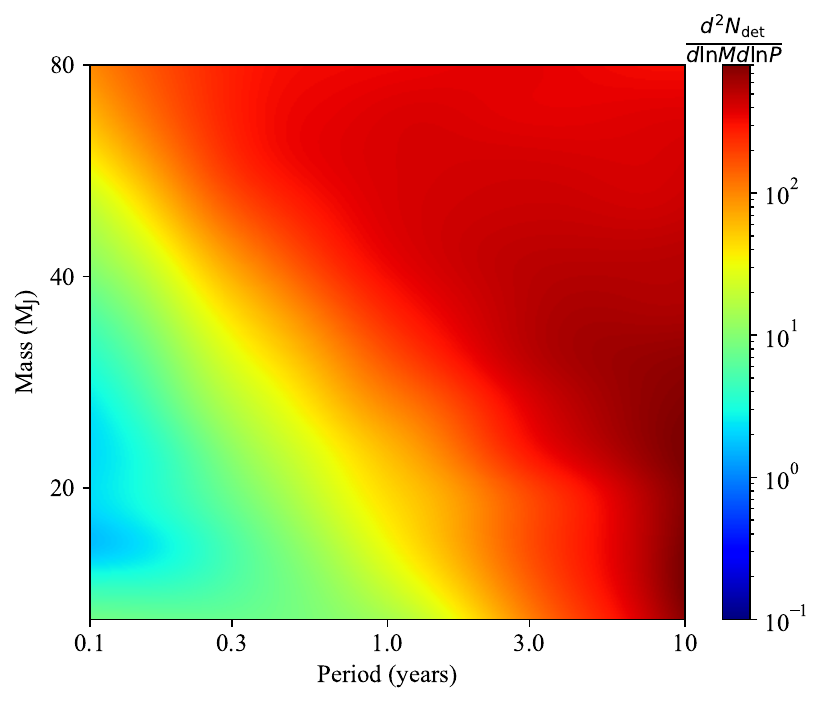}}
    \caption{DR4 brown dwarf yield from this sample assuming distributions from \citet{lineweaver06} and our study with DR3.}
    \label{fig:bd_yield}
\end{figure}

\noindent In order to predict the possible brown dwarf detection with DR4, we assumed a distribution of companions based on previous studies, and combined this with the detection probability.  Initially, we assume a broken power-law distribution of brown dwarf mass given by \citet{lineweaver06} in which
\begin{equation}
    \frac{dN}{d\mathrm{ln}Md\mathrm{ln}P}\propto M^{\alpha}P^{\beta},
\end{equation}
where $\alpha$=-9.4 for M$<44$\,M$_{\mathrm{J}}$, $\alpha$=23.1 for M$>44$\,M$_{\mathrm{J}}$ and $\beta$=0.3.  The distribution is normalised such that the total number of brown dwarf mass companions is 0.005, as the proportion of stars with brown dwarf companions was found to be 0.5\,\%.  For each source in our sample, we multiplied this distribution by the detection probability to get the expected number of detectable brown dwarfs as a function of mass and period.  This method was repeated using the distribution calculated in Section~\ref{sec:inference}.  Both resultant brown dwarf yields are shown in Figure~\ref{fig:bd_yield}.

\noindent As expected, comparing the expected yields from the conservative distribution in \citet{lineweaver06} with our more optimistic estimate produces very different results.  Integrating these distributions, with the \citet{lineweaver06} distribution, we can expect $\sim$134 detectable brown dwarfs with periods less than 5\,years, corresponding to 0.12\,\% of the sample having a detectable brown dwarf companion in this period range.  Assuming the distribution from Section~\ref{sec:inference}, we can expect 1327 detectable brown dwarfs in this range, corresponding to 1.2\,\% of the sample.\\
\noindent These vastly different results show that, even with the conservative distribution, we can expect to detect a sample of companions in the brown dwarf desert with DR4 and, with epoch data, we will be able to obtain accurate mass estimates.
\section{Summary and Conclusions}
In this study, we have investigated the presence of companions to Sun-like stars in nearby systems using the limited information available in DR3.  From this, we can conclude that \textit{Gaia} is highly sensitive to brown dwarf mass companions on short (2-4\,year) periods which has made it an ideal tool for exploring the brown dwarf desert.  We have identified several sources which could be harbouring a brown dwarf companion in this period range but are limited by degeneracies in which different companion properties can produce the same RUWE value.\\
\noindent Assuming high RUWE can be explained by the presence of one companion, we ran Bayesian Inference to determine the companion properties and discovered a significant percentage with possible brown dwarf companions.  Accounting for observational biases, we inferred an increase in brown dwarf abundance as a function of period up to $\sim$1\,year, a flat distribution from 1--10\,years, and a decrease for longer periods.  We see the same behaviour when comparing two separate populations of brown dwarfs, above and below 42.5\,M$_{\mathrm{J}}$.  Similar behaviour is displayed for low-mass stars.  Our results suggest brown dwarf companions are more abundant than previously implied, due to \textit{Gaia}'s high sensitivity in this mass-period range, when compared to previous RV studies.\\
\noindent However, if we allow for two companions, we find that the results favour planetary mass companions and brown dwarfs at the lower mass range.  Relying purely on the RUWE of DR3, we demonstrate the source of degeneracy in our results.  This high uncertainty means that, contrary to our earlier result, the brown dwarf desert may be `drier' than initially implied when we assumed one companion.  Figure~\ref{fig:epoch_ruwe} demonstrates the wide range of epoch data which can produce similar RUWE values.  With DR4, it will be possible to remove this degeneracy and properly probe the brown dwarf desert.\\
\noindent In this study, we have shown the region in mass-period space in which \textit{Gaia} DR3 is the most sensitive, and how this aligns with the brown dwarf desert.  For DR4, the peak sensitivity will shift to longer periods ($\sim$5\,years) but with more data, it will be possible to observe multiple orbits of companions on shorter periods.  Applying prior knowledge of brown dwarf distributions, we demonstrated the expected high brown dwarf yield from DR4.  With more data, DR4 will also contain more accurate measurements of stellar parallax and proper motion.  This has the effect of reducing the threshold RUWE for detectable companions, thus providing a larger sample in which to search for brown dwarfs.  Combined with the extra information in epoch astrometry, this will allow a full statistical study to efficiently map the brown dwarf desert in detail for the first time.
\begin{acknowledgement}
This work was funded by the Australian Research Council Discovery Grant DP210100018.\\
This work presents results from the European Space Agency (ESA) space mission Gaia. Gaia data are being processed by the Gaia Data Processing and Analysis Consortium (DPAC). Funding for the DPAC is provided by national institutions, in particular the institutions participating in the Gaia MultiLateral Agreement (MLA). The Gaia mission website is \newline\href{https://www.cosmos.esa.int/gaia}{https://www.cosmos.esa.int/gaia}.\newline The Gaia archive website is \href{https://archives.esac.esa.int/gaia}{https://archives.esac.esa.int/gaia}.\\
The Flatiron Institute is a division of the Simons Foundation.
\end{acknowledgement}

\section*{Data Availability Statement}
The data underlying this article are available from the corresponding author on reasonable request.  The inference program and the data file are  available at \newline\href{https://https://github.com/awallace142857/brown_dwarf_gaia}{https://github.com/awallace142857/brown\_dwarf\_gaia}.

\printendnotes
\begin{appendix}
\section{Testing Inference Methods}
\subsection{Injection and Recovery}
\label{sec:test_inference}
\begin{figure}[t]
    \subfigure[30\,M$_{\mathrm{J}}$ brown dwarf on 0.5\,year orbit]{\label{fig:30_0.5} \includegraphics[width=0.8\linewidth]{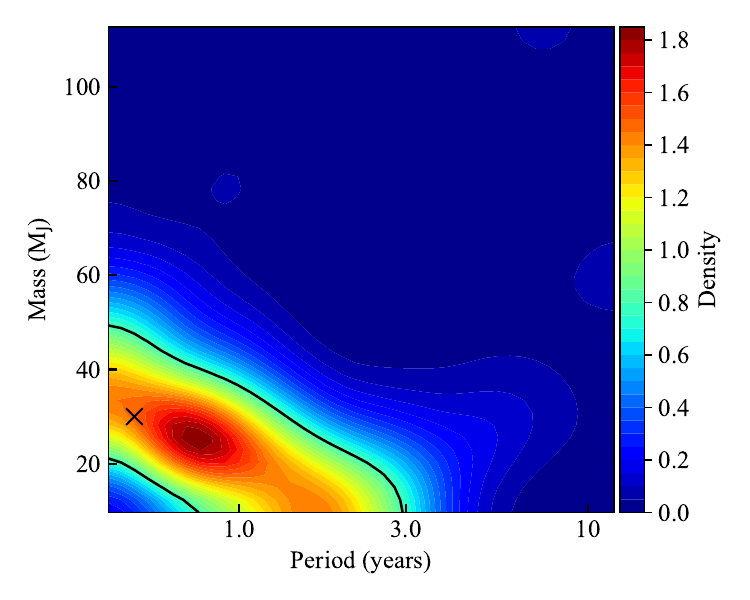}}
    \subfigure[80\,M$_{\mathrm{J}}$ star on 2\,year orbit]{\label{fig:mass_dist} \includegraphics[width=0.8\linewidth]{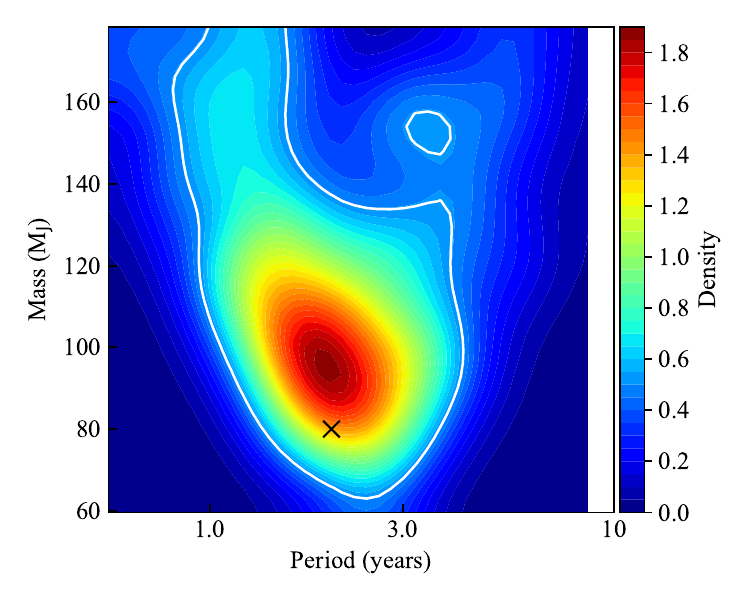}}
    \subfigure[50\,M$_{\mathrm{J}}$ brown dwarf on 4\,year orbit]{\label{fig:mass_dist} \includegraphics[width=0.8\linewidth]{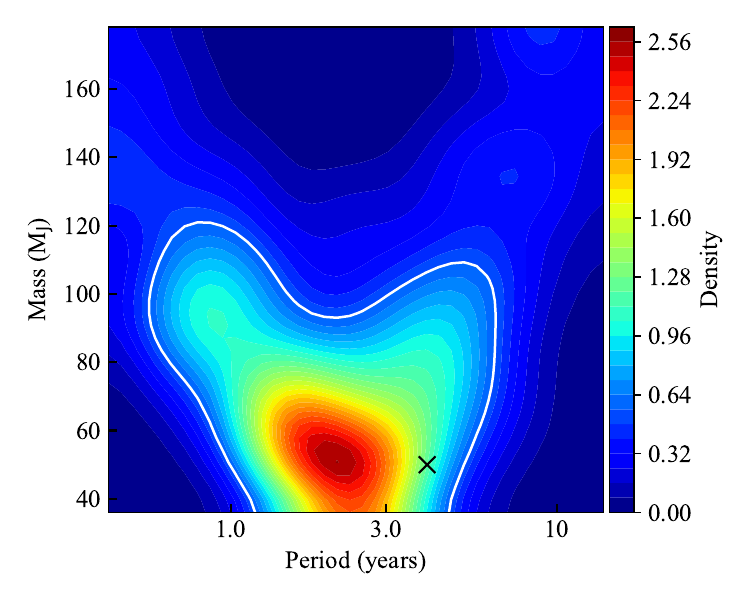}}
    \caption{Example posteriors of mass and period from simulated data.  Injected values are shown with crosses and 1-$\sigma$ levels with contours.  Our solutions seem to favour periods close to 2.8\,years which should be taken into account when computing occurrence rates in Section~\ref{sec:inference}.}
    \label{fig:test_cases}
\end{figure}

The Bayesian inference of companion mass and period is shown for three injected examples for a star at a distance of 100\,pc, using simulated data as the input.  The injected values of mass and period are compared to posteriors in Figure~\ref{fig:test_cases}.  Each companion has an eccentricity of 0.1, $i$, $\Omega$ and $\omega$ of 0$^{\circ}$ and $T_{P}$ of 2016.0.  The injected values are shown with crosses and 1-$\sigma$ levels are also indicated.
\subsection{Using a Nearby Reference Star}
\label{sec:ref_ex}
\begin{figure}[t]
    \centering
    \includegraphics[width=1.0\linewidth]{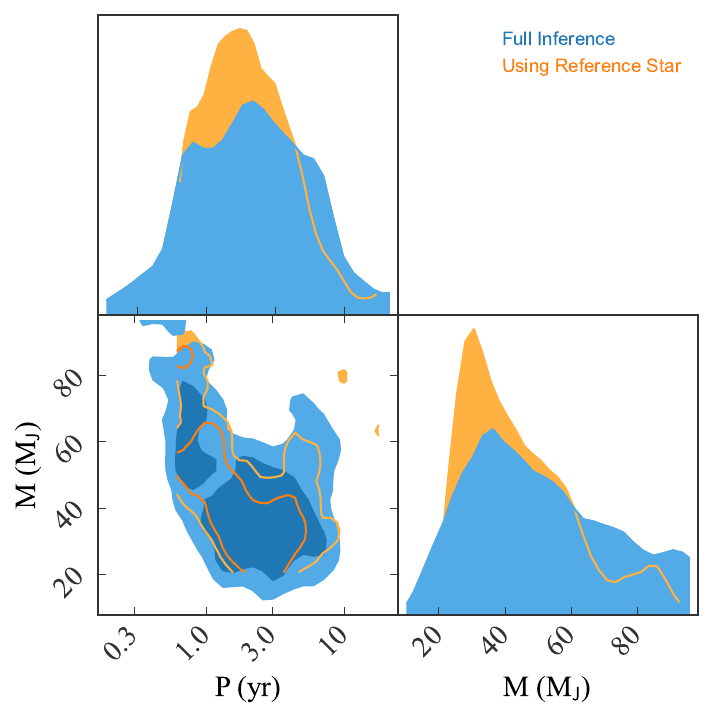}
    \caption{Posterior distributions of inferred mass and period of a companion to \textit{Gaia} DR3 7534235926299136 using direct inference and using \textit{Gaia} DR3 10176053130242048 as a reference source.}
    \label{fig:ref_ex}
\end{figure}

Figure~\ref{fig:ref_ex} shows posteriors for an example distant source \textit{Gaia} DR3 7534235926299136 with a parallax of 5.69\,mas.  This source is paired with \textit{Gaia} DR3 10176053130242048 which has a parallax of 11.7\,mas.  These sources are separated by 2.89\,$^{\circ}$.  Figure~\ref{fig:ref_ex} shows the posterior distributions of \textit{Gaia} DR3 7534235926299136 by direct inference (blue) and using the reference source and correcting for distance, mass, error and RUWE.
Figure~\ref{fig:ref_ex} shows similarly shaped mass and period posteriors for the two methods of calculation.  The companion masses from direct inference and using a reference star are 46.9$^{+28.8}_{-18.4}$\,M$_{\mathrm{J}}$ and 40.5$^{+20.2}_{-12.4}$\,M$_{\mathrm{J}}$ respectively and the periods are 2.13$^{+3.74}_{-1.39}$\,years and 1.90$^{+2.26}_{-0.95}$\,years respectively.  These values are close enough within uncertainty that we have confidence in this approach for the more distant sources, reducing the total processing time.
\end{appendix}
\bibliography{references}
\end{document}